\documentclass{aa}  %
\usepackage{graphicx}
\usepackage{aalongtable,lscape}
\usepackage{natbib}
\bibpunct{(}{)}{;}{a}{}{,}
\usepackage{epsfig}
\usepackage{graphicx}
\usepackage{lscape}
\usepackage{setspace}
\usepackage{multirow}
\usepackage[tbtags,fleqn]{amsmath}

\begin{document}
  \title{Strong Lensing Analysis of the Cluster RCS0224-0002 at $z=0.77$}
  \subtitle{}
  \author {J. Rzepecki\inst{1}\and
    M. Lombardi\inst{1} \and
    P. Rosati\inst{1} \and
    A. Bignamini\inst{2} \and
    P. Tozzi\inst{2}
  }
\authorrunning{J.\ Rzepecki et al.}
\titlerunning{Strong Lensing Analysis of the Cluster RCS0224-0002 at $z=0.77$}
\offprints{J.\ Rzepecki}
\institute{European Southern Observatory, Karl-Schwarzschild-Strasse 2,85748 Garching bei Munich, Germany\\
  \email{jrzepeck@eso.org}
  \and
   INAF, Osservatorio Astronomico di Trieste, via G.B. Tiepolo 11, 34131, Trieste, Italy\\
}
\date{}
\abstract
    {}
    {We present a detailed mass reconstruction of the cluster RCS0224-0002 at $z=0.773$ from the strong lensing features observed with  HST/WFPC2.}
    {The mass profile is reconstructed using a parametric approach. We introduce a novel method to fit extended multiple images based on the Modified Hausdorff Distance between observed arcs and the arcs reproduced by the model. We perform the detailed error analysis of the model parameter using the MCMC method.}
    {Our model reproduces all the observed strong lensing features of the RCS0224-0002 and predicts the redshift of one of the arcs systems to be $z\approx 2.65$ (the other system has an spectroscopic redshift of $z=4.87$). The reconstructed inner mass profile is well fitted  by a non-singular isothermal sphere, rather than with an NFW model. Dark matter substructure, derived from the light distribution of the most luminous cluster members, is crucial for reproducing the complexity of the quadrupole image system, which could not be achieved otherwise. The reconstructed mass distribution closely follows the light, however it is significantly shifted from the X-ray emission of the gas. The mass of RCS0224-0002 derived from the lensing model, $\approx 2\times10^{14}\, M_\odot$ is in a very good agreement with the one obtained from the X-ray temperature measured with deep Chandra observations.}
    {}
    \maketitle
%

\section{Introduction}
\label{sec:introduction}

The presence of dark matter is evident in galaxy clusters, where the
mass to light ratios, $M/L\approx 200\ h\ (\mathrm{M_\odot/L_\odot})$,
by far exceeds the mass of stars, and the hot gas implied by X-ray
data \citep{hradecky}. Gravitational lensing is one of the most
attractive methods to directly study the mass distribution in the
Universe on different scales, regardless of its component and
dynamical state (see e.g. {\it ESA-ESO Working Groups, Report No. 3},
\citealp{ESOESA}).  Strong gravitational lensing in galaxy clusters
leads to the formation of multiple images and giant luminous arcs
\citep[e.g.][]{schneider}.  Observations of these features allow us to
investigate the distribution of the mass responsible for the
deflection, and in particular provide accurate estimates of the
\textit{total\/} mass within giant arcs and the innermost density
profile, which can be compared with N-body simulations.  In addition,
mass models can reveal the location of highly magnified background
galaxies and allow spectroscopic studies of very distant faint, highly
magnified sources, that would be under normal circumstances beyond the
reach of 10m-class telescopes.

Accurate estimates of the mass profiles of galaxy clusters are
fundamental for the modern cosmology since they provide severe tests
for the theories of structure formation.  The most popular theory
based on the assumption of non interacting cold dark matter predicts a
universal profile (NFW, \citealp{nfw}) rather then an isothermal
profile.  For the lensing cluster Abell 1689, \citet{broadhurst} ruled
out the isothermal profile with 10$\sigma$ confidence.  However,
another study of the same cluster by \citet{halkola} shows
that both the elliptical NFW and the isothermal softened elliptical
fit the data well.  Hence, although deep observations of this cluster
showed a formidable arc system, \textit{parametric\/} strong lensing
models surprisingly lead to different mass density profiles,
highlighting the difficulty of current inversion techniques in
determining the uniqueness of the solution and the real uncertainties
of the reconstructed mass maps

Here we present a study of the cluster RCS0224-0002 at $z=0.773$ which
was discovered as a part of the Red-Sequence Cluster Survey (RCS,
\citealp{gladders}).  After the identification of the main strong
lensing features of this cluster with VLT spectroscopy, follow-up
observations were carried out with HST-WFPC2 by \citet{gladders}, in
X-rays with the Chandra observatory \citep{hicks}, and in sub-mm using
SCUBA on the JCMT \citep{webb}.

We construct a parametric model of the projected mass density
distribution of RCS0224-0002 based on its strong lensing features, one
of which with secure redshift.  The method used in this paper to
construct the best mass model is based on the so-called Modified
Hausdorff Distance (MHD, \citealp{dubuisson}), and has the advantage of allowing
us to use the information provided by the sub-arcsecond morphology of
arcs.  We compare the mass distribution with the spatial distribution
of the hot gas obtained form the X-ray data.  

When we were finalizing this paper, a lensing model of the same
cluster has been independently presented by \citet{swinbank}.  However,
these authors focus their work on the properties of a highly magnified
$z=4.87$ galaxy observed in the field; moreover, their lensing model,
which is based only on the constraints provided by a single arc system
(the giant arc labeled A in Fig.~\ref{fig.arcs}), is significantly
different from ours.

The paper is structured as follows: In Sect.~\ref{Observations} we
present data available on the RCS0224-0002. In
Sect.~\ref{Arc_identification_and_cluster_members} we discuss the
strong lensing features and the red sequence of the RCS0224-0002.
Section \ref{X-ray emission} is dedicated to the X-ray emission of the
RCS0224-0002.  In Sect.~\ref{Model} we present assumptions behind our
model and the method we use to obtain the projected mass
distribution. In Sect.~\ref{Results} we present and discuss our
results. In Sect.~\ref{error} we perform error estimation. And finally
in Sect.~\ref{conclusions} we present our conclusions.

In this paper we use a standard cosmological model with
$\Omega_\mathrm{m}=0.3$, $\Omega_\mathrm{\Lambda}=0.7$, and
$\mathrm{H_0}=72\ \mathrm{km}\ \mathrm{s}^{-1}\ \mathrm{Mpc}^{-1}$. We
give all the magnitudes in the AB system, if not otherwise specified.

\section{Observations}
\label{Observations}

The HST observations of the RCS0224-0002 were taken on the 2001/08/20
in two filters, F606W and F814W using the WFPC2 camera (PI: Gladders,
Proposal ID: 9135). The target coordinates were RA: 02:24:30.82, DEC:
$-$00:02:27.8 and the exposure time for each filter was 1100
seconds. The WFPC2 data reduction was performed by Associations
Science Products Pipeline.\footnote{{\it http://archive.eso.org/archive/hst/wfpc2\_asn/wfpc2\_products.html}}

The X-ray data were taken on the 2002/11/15 with the ACIS-S instrument
on the Chandra observatory (PI: Gladders, Proposal Num: 03800013). The
target coordinates were RA: 02:24:34.10, DEC: $-$00:02:30.90 and the
exposure time was 14560 seconds. On the 2004/12/09, RCS0224-0002 was
observed with the ACIS-S again (PI: Ellingson, Proposal ID: 05800899)
with exposure time of 90150 seconds. The two ACIS-S observations were
combined with CIAO 3.3, using CALDB 3.2.1, leading to 100.8 ksec of
effective exposure time. Details on the reduction and spectral
analysis, whose results are given below, can be found in \citet{balestra2007}.

\section{Arc identification and cluster members}
\label{Arc_identification_and_cluster_members}

\begin{figure}
   \centering
   \includegraphics[scale=0.4]{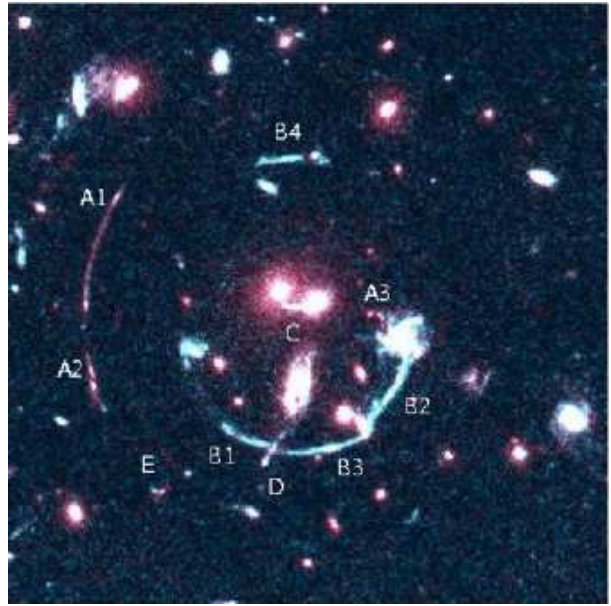}
   \caption{The RCS0224-0002 cluster with labeled arcs. Color image composed from F814W and F606W WFPC2 HST images. The image is 40 arcsec across.}
   \label{fig.arcs}
 \end{figure}

RCS0224-0002 has seven prominent luminous arcs and arclets marked as
A1, A2, A3, B1, B2, B3, and B4 in Fig.~\ref{fig.arcs}. Unfortunately,
out of those seven arcs, only one arc system (A) has a confirmed
spectroscopic redshift of 4.87 \citep{gladders}. The same authors
estimated the redshift of system B within the range 1.4 to 2.7 based
on the lack of emission lines in their spectra.  Since the redshifts
of arcs B1, B2, B3, and B4 are not known, an assumption needs to be
made of whether all those arcs are images of one source or more
sources.  Based on very similar color, structure and distance from the
center of the cluster we suppose that arcs B1, B2, B3, and B4 are
images of one source and we call it system B.  This conjecture is
supported by the lensing model described below, since by assuming the
existence of two separate systems (B1--B2, B3--B4) our model predicts
relatively bright multiple images which are not observed.  We excluded
that the feature D is a radial arc, despite its elongated morphology,
since no tangential counter images are visible and because its
position and morphology makes this hypothesis unlikely.  Our model
suggests that feature C is a central demagnified image, which is clearly
visible in Fig.~\ref{fig.cd_subtracted} showing the F606W image
after subtracting the two cD galaxies. There is also a very faint red
arc, labeled E, which was not included in our analysis.

\begin{figure}
   \centering
   \includegraphics[scale=0.3]{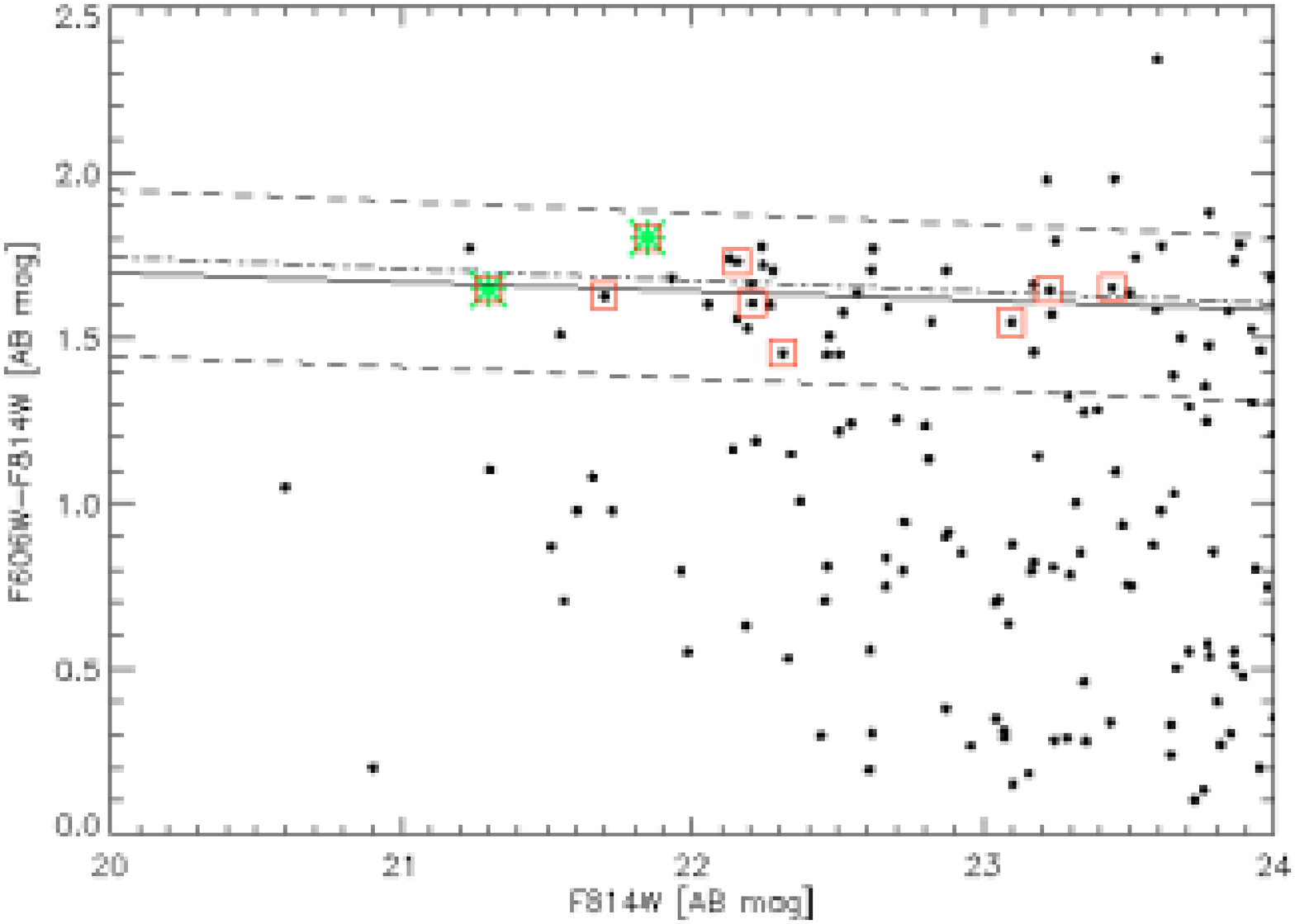}
   \caption{ The color-magnitude diagram of RCS0224 with the WFPC2
     F606W/F814W filters. The dots represent all objects in the
     field.  The squares represent the cluster red sequence (galaxies
     within 15 arcsec from the cluster center), the stars mark two
     central galaxies. The solid and dot-dashed lines are our best fit
     to the red sequence and the one of MS1054 at similar redshift. }
   \label{fig.color-mag}
 \end{figure}

Since mass is known to follow light in galaxy clusters \citep[see
e.g.][]{sand}, the distribution of color selected cluster members is
often used to model substructure of the underlying dark matter.
Besides to the two brightest central galaxies (BCGs), there is no
public spectroscopic information available in the field, we then used
the red sequence to identify likely cluster members. In
Fig.~\ref{fig.color-mag} we show the color-magnitude diagram over the
whole WFPC2 field, highlighting red sequence objects lying withing
15\arcsec\ from the cluster core.  Photometry was performed using
SExtractor software \citep{bertin}, by detecting sources in the F814W
band and measuring F606W${} - {}$F814W colors with aperture of 1''
diameter\footnote{The WFPC2 zero points were calculated according to:
  $ZP_{AB}=-2.5\log{(PHOTFLEM)}-21.1-5\log{(PHOTPLAM)}+18.6921$}. The
solid and dot-dashed lines represent our best fit to the red sequence
and the best fit found by \citet{best2002} for the cluster MS1054 at
$z=083$ for the same filters, after applying a K-correction of 0.07
mag. Red sequence objects were defined as those within $\pm 0.25$ mag
of the best fit line.

\begin{figure}
   \centering
   \includegraphics[scale=1.0]{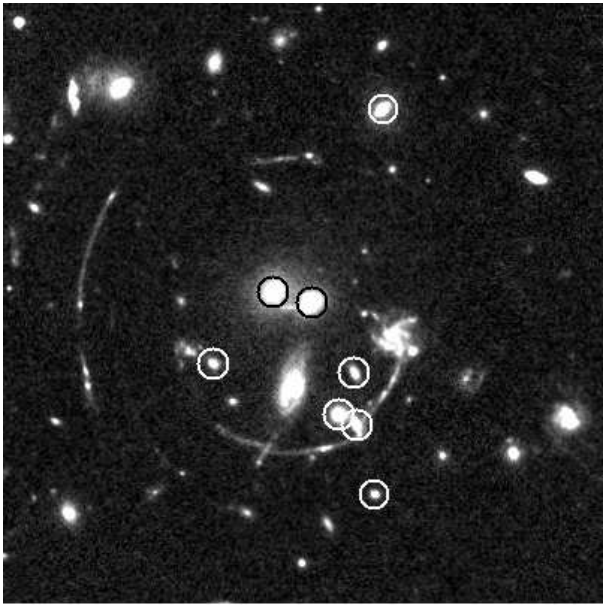}
   \caption{The red sequence galaxies visible on the F814W filter
     WFPC2 HST image (objects that are marked here correspond to the
     squares in Fig \ref{fig.color-mag}.}  \label{fig.redseq}
 \end{figure} \begin{figure} \centering
   \includegraphics[scale=0.4]{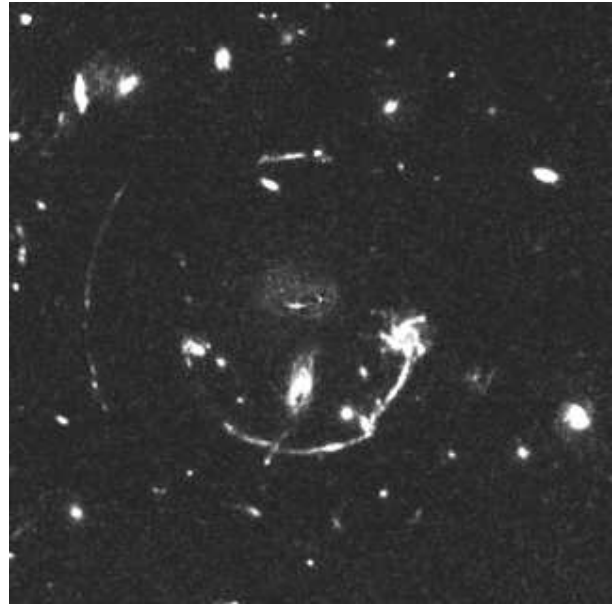}
   \caption{RCS0224-0002 in the F606W filter with subtracted cD
     galaxies. The central radial feature C is clearly visible.}
   \label{fig.cd_subtracted} \end{figure} 
\section{X-ray emission}
\label{X-ray emission} 
\begin{figure} \centering
   \includegraphics[scale=0.4]{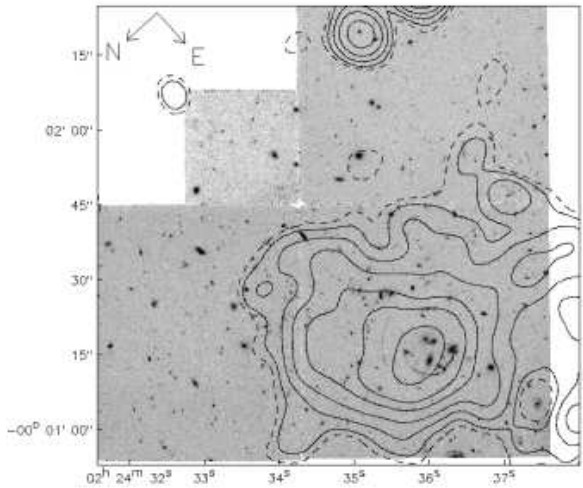}
   \caption{The X-ray emission contours of RCS0224-0002 (smoothed with
     a Gaussian with $\sigma=5\arcsec$) over-plotted on the F606W
     WFPC2 HST image. }  \label{fig.xray} \end{figure}

\begin{figure}
   \centering \includegraphics[scale=0.4,
   angle=-90]{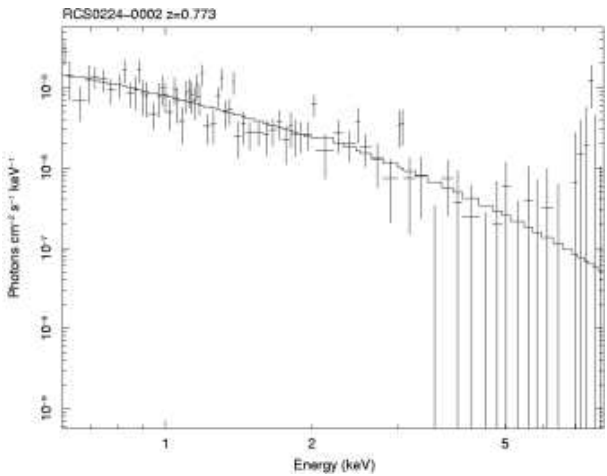}
   \caption{X-ray spectrum of RCS0224-0002 from 100ksec Chandra
     observations, with the best fit Mekal model, for $kT =
     5.26_{-1.07}^{+1.14}$ keV. }  \label{fig.xray-spec} \end{figure}

The X-ray emission traces the hot gas trapped
in the cluster potential well.  The gas itself contributes about 15\%
to the total mass of the cluster and for relaxed systems traces
closely the total mass density distribution. We overlay the X-ray
contours of RCS0224-0002 from the 100 ksec Chandra observations in
the 0.5--2 keV band onto the WFPC2 image in Fig.~\ref{fig.xray}.  The
overall X-ray emission is not symmetric, with a plume extending NW,
and its peak shifted $\sim\!5$ arc seconds north from the two central
BCGs. To measure the X-ray temperature, we used and extraction region
of 36.7 arcsec (or 265 kpc), which encompasses most of the X-ray
emission by maximizing the signal-to-noise.  The background
subtracted, unfolded spectrum is shown in Fig.~\ref{fig.xray-spec}.
We used Xspec v.12.3.0 \citet{arnaud1996} to fit the data with a
single temperature Mekal model \citep{kaastra1992, liedahl1995} and
model the Galactic absorption with tbabs \citep{wilms2000}, fixing
the Galactic neutral Hydrogen columns density to the Galactic value
obtained with radio data \citep{dickey1990}.  Since the
signal--to--noise ratio in each energy bin is low, we used the
C-statistics for the best fit model, over the energy range 0.6-8.0
keV (excluding low energy photons due to uncertainties of ACIS
calibration). We used $742\pm 35$ total net counts in the fit
($514\pm 23$ in the soft 0.5-2 keV band) and found a best fit
temperature of $kT = 5.26_{-1.07}^{+1.14}$ keV (1-sigma error). The
de-absorbed flux within the extraction aperture, in the (0.5 - 2.0)
keV band, is $1.84\times 10^{-14}$erg cm$^{-2}$ s$^{-1}$ and the
rest-frame X-ray luminosity $L_X (0.5-2 \rm{keV}) = (0.38\pm 0.02)
\times 10^{44}$ erg s$^{-1}$. The bolometric luminosity returned by
the best fit model is $L_{BOL} = (1.28\pm 0.06) \times 10^{44}$.
With these values of X-ray luminosity and temperature, we note that
RCS0224-0002, which is an optically selected cluster, lies on the
$L_X-T$ relation determined from large samples of X-ray selected
clusters \citep[e.  g.][]{rosati2002} 

We can use the measured cluster
temperature to estimate the cluster mass assuming the hydrostatic equilibrium and isothermal distribution of the gas, with a polytropic index $\gamma=1$. Using the standard $\beta$-model for the gas density profile, $\rho_{\rm gas}(r)=\rho_0/[1+(r/r_c)^2]^{3\beta /2}$, the mass within the radius $r$ can be written as \citep{sarazin}:

\newcommand{\hm}{\,h^{-1}{\rm Mpc}} 
\newcommand{\msun}{\,h^{-1}M_\odot}
\begin{align}
  M(<r)\simeq 1.11 \times 10^{14}\beta \gamma {T(r)\over {\rm
    keV}}{r\over \hm}{(r/r_c)^2\over 1+(r/r_c)^2}\msun\,,
\label{eq:begam}
\end{align}
A fit to the X-ray surface brightness profile with the corresponding $\beta$-model $\mathrm{SB}(r)\propto[1+(r/r_c)^2]^{-3\beta+1/2}$ yields a core radius $r_c = (253\pm 72)\mathrm{kpc}$ and $\beta = 0.97\pm 0.3$. Therefore the mass within $R_{200}=0.4\ \mathrm{Mpc}$ is $(1.7\pm1.1)\times 10^{14}M_\odot$.

 
\section{Model} 
\label{Model} 
We constructed the mass model of
RCS0224-0002 by fitting the position and shapes of the multiple image
systems A, B and C. Based on the light distribution of most luminous
red-sequence galaxies, our model consists of several mass components:
two isothermal non-singular ellipsoids to reproduce global cluster
properties (NIE1, NIE2); eight isothermal non-singular spheres fixed
at the position of cluster members (NIS1..8) - refereed to as the
substructure; one non-singular ellipsoid, corresponding to the
elongated object marked D in Fig.~1 (NIE3).  In order to reduce the
number of free parameters, we fixed the positions and the
\textit{relative\/} masses of the galaxy cluster clumps using the
optical data available.  In summary, we have 17 adjustable parameters
in our model, including sources positions and unknown redshifts.  All
parameters are listed in Tables \ref{table:1_0} and \ref{table:1_1}.
The seven observed extended images are enough to constrain those 17
parameters due to the fact that we base our goodness of fit function
not only on the position of the images but on the full information
encoded in their shapes.  Models including the radial feature D as a
counter-image give the worst results, but as mentioned in
Sect.~\ref{Arc_identification_and_cluster_members}, it is probably an
foreground edge-on galaxy. Arc E was not used in the model since its
redshift is unknown and it is too faint to provide any further
constraint. We would like to emphasize that we do not assign any
physical meaning to the two distinct smooth components (NIE1\&NIE2),
and we are interested in the properties of the overall, combined
profile. We have also tried to fit the data with only one smooth
component (NIE1) and the substructure, however in that case we were
not able to fit the arcs system B accurately.  
\subsection{Mass profiles} 
Although the N-body simulations of dark matter halo
formation suggest NFW profiles rather than isothermal ones, recent
strong lensing studies do not exclude and in some cases even prefer
isothermal profile over NFW \citep{gavazzi, halkola}. We model here
all mass components as non-singular isothermal ellipsoid, a simple
generalization of a non-singular isothermal sphere often used as a physical
representation of a gravitationally relaxed system.  The use of
isothermal profiles has also the advantage of being computationally
less demanding. The associated gravitational potential $\phi$,
projected mass density $\rho$, and deflection angle $\vec{\alpha}$
are given by 
\begin{align}
   \phi(x_1,x_2) \equiv{}&  x_1\frac{\partial\phi}{\partial x_1} + x_2\frac{\partial\phi}{\partial x_2}\label{eqn.iso_ell:1}\\
   &{} - b q s\ln\left[(\psi+s)^2 + (1-q^2)x_1^2\right]^{1/2} \nonumber\\
   &{} + b q s \ln\left[(1+q)s\right]\; , \nonumber\\
   \alpha_1(x_1,x_2) \equiv{}& \frac{\partial\phi}{\partial x_1} = \frac{bq}{\sqrt{1-q^2}}\mathrm{atan}\left[\frac{\sqrt{1-q^2}x_1}{\psi+s}\right]\; ,\label{eqn.iso_ell:2}\\
   \alpha_2(x_1,x_2) \equiv{}& \frac{\partial\phi}{\partial x_2} = \frac{bq}{\sqrt{1-q^2}}\mathrm{atanh}\left[\frac{\sqrt{1-q^2}x_2}{\psi+q^2s}\right]\; ,\label{eqn.iso_ell:3}\\
   \psi^2(x_1,x_2) \equiv{}& q^2(s^2+x_1^2)+x_2^2\; ,\\
   \rho(x_1,x_2) \equiv{}&
   \frac{1}{2}\frac{b}{(s^2+x_1^2+x_2^2/q^2)^\frac{1}{2}}\; ,
\end{align} 
where $q$ is the ellipticity, $s$ is the core radius, $b$ is the scale
factor and $\vec{x}=(x_1,x_2)$ is the position in the image plane.
Note that the fast method for calculating the deflection angle of the
softened non-singular ellipsoid might be found in \citet{barkana1998}.
\subsection{Minimization method} 
\subsubsection{Source plane minimization} 
\label{sourceplanemin} 
In order the get a first, approximated solution, we perform model
fitting minimization on the source plane.  This technique is
computationally very efficient, since there is no need to solve the
inverse problem of the lensing equation and the deflection angle is
only computed at the position of the images. We also assume that
sources are small compared to the scale of variations of the lensing
potential.  If we have $N$ images at positions $\{ \vec x_\mathrm{i}
\}$ corresponding to one source, then we define the $\chi^2$ as
\begin{align}
   \label{eq:8} \chi^2_{\mathrm{src}} {} = \sum\limits_i\delta \vec
   u_i^\mathrm{T}\mu^{\mathrm{T}}_iS_i^{-1}\mu_i\delta \vec u_i + P \;
   , \end{align} where \begin{align} \label{eq:9}
   \delta \vec u_i ={}& \vec u_{\mathrm{obs}, i}-\vec u_{\mathrm{mod}} \; ,\\
   \label{eq:10} \vec u_{\mathrm{obs},i} ={}& \vec x_{\mathrm{obs},i}
   - w\nabla\phi(\vec x_{\mathrm{obs},i}) \; .  
\end{align} 
In the previous equations, $\vec u_{\mathrm{obs},i}$ is the source position
(as predicted by the model) corresponding to the image $\vec
x_{\mathrm{obs},i}$, $\phi({\vec x}_{\mathrm{obs},i})$ is the lensing
potential at image $i$, $w$ is the cosmological weight of the source
\citep[see e.g.][]{lombardi}, and $\mu_i$ is the magnification matrix
(inverse of the Jacobian matrix of the lens mapping) at the image
$i$.  Moreover, in Eq.~\eqref{eq:8} we introduced the covariance
matrix of the position measurements $S$.  For simplicity, in this
paper we assume that the covariance matrix is diagonal and takes the
form 
\begin{align} \label{eq:11} S_i {} = \left[ \begin{array}{cc}
       \sigma_i^2 & 0 \\
       0 & \sigma_i^2 ,\\
     \end{array} \right] = \sigma_i^2I \; , 
\end{align} 
where $\sigma_i$ is estimated to be ${} \sim 0.05''$.  In the definition of
our $\chi^2$ [Eq.~\eqref{eq:8}] we introduced also a ``penalty''
function $P$.  This function, is used to bound some of the free
parameters to certain intervals, and is chosen to have the functional
form 
\begin{align}
   P = {} & \mathcal{P} \sum\limits_{p=0}^{N}\mathrm{atan} \bigl(10^7 (b_{\mathrm{down},i}- p_i) \bigr) \\
   &{} + \mathrm{atan}(10^7(p_i - b_{\mathrm{up},i})) + \mathcal{P}\pi
   \; ,\notag 
\end{align} 
where, $N$ is the number of bounded parameters in our model, $p_i$ is the $i$-th bounded parameter, which
is required to be in the range $[b_{\mathrm{down},i},
b_{\mathrm{up},i}]$.  Note that the penalty function $P$ behaves
similarly to a ``square potential well'', i.e. the sum of two
Heaviside functions; however, the use of analytic functions ensures
that $P$ is differentiable and makes our minimization numerically
stable.  In order to effectively bound our parameters, we used a
large number for the coefficient $\mathcal{P}$.  When there is more
then one source, the same procedure is repeated for all sources and
the resulting $\chi^2$ from Eq.~\eqref{eq:8} are added.  The
magnification matrix $\mu_i$ is included because $\mu_i\delta\vec
u_i\approx \delta \vec x_i$, so that $\chi_{\mathrm{src}}^2$ is an
approximation of $\chi^2$ in the image plane. However, this also
introduces a weight in the $\chi^2$ term, as images for which
$\mu_i\delta\vec u_i$ are small do not contribute significantly to
the minimization process.  It is possible to write an analytical
expression for the source position
that minimizes $\chi^2_{\mathrm{src}}$:\\

\begin{align}
   \vec u_{\mathrm{mod}} ={}& A^{-1}\vec b; ,\\
   A ={}& \sum\limits_i\mu_i^TS_i^{-1}\mu_i\; ,\\
   \vec b ={}& \sum\limits_i\mu_i^TS_i^{-1}\mu_i\vec
   u_{\mathrm{obs},i}\;.  
\end{align} 
\subsection{Extended images} 
The best fit model provided by Eq.~(\ref{eq:8}) is used as starting point
for the image plane analysis.  This step is based on a new $\chi^2$
minimization, with a $\chi^2$ composed of two terms: the so-called
Modified Hausdorff Distance (MHD, \citealp{dubuisson}) between the
modeled and observed image sets and the ``plain difference'' between
the same sets.  The MHD between two sets $A$ and $B$ is defined as
\begin{align}
   \mathrm{MHD} ={}& \max(h_{\mathrm{ab}}, h_{\mathrm{ba}})\;,\label{mhd}\\
   h_{\mathrm{ab}} ={}& \frac{1}{\| A\|}\sum\limits_{a\in A}\min\limits_{b\in B}\| a-b\|^2\;,\\
   h_{\mathrm{ba}} ={}& \frac{1}{\| B\|}\sum\limits_{b\in
     B}\min\limits_{a\in A}\| a-b\|^2\;.  
\end{align} 
In addition, the
``plain difference'' between the observed and modeled arcs is
computed as follows. All pixels in each observed arc system,
generically called $O$, are assigned a value of $1$; other pixels are
assigned a value of $-1$. The same procedure is applied to the
corresponding modeled arcs ($M$) and the difference
$\mathrm{diff}(O,M) = |O - M|$ is calculated.  In summary, the
expression to minimize in the image plane is \begin{align} X^2 =
\mathrm{MDH}(D,M)+\omega\ \mathrm{diff}(M,D) + P\;.  \end{align}
The factor $\omega$ was chosen to be $\sim 0.1$, since this value
resulted in the fastest convergence.  The penalty function $P$ is
used to bound some of the model parameters and it is defined in
Sect.~\ref{sourceplanemin}.  By using two distance components, we
ensure an efficient convergence of the minimization since when the
modeled and the observed images start to overlap, the MHD becomes
less sensitive to small variations then the plain difference.  The
{\it Powell} algorithm \citep{powell} is used for all the
minimization procedures.  
\section{Results} 
\label{Results} 
\begin{figure}[p]
   \centering
   \includegraphics[scale=0.6]{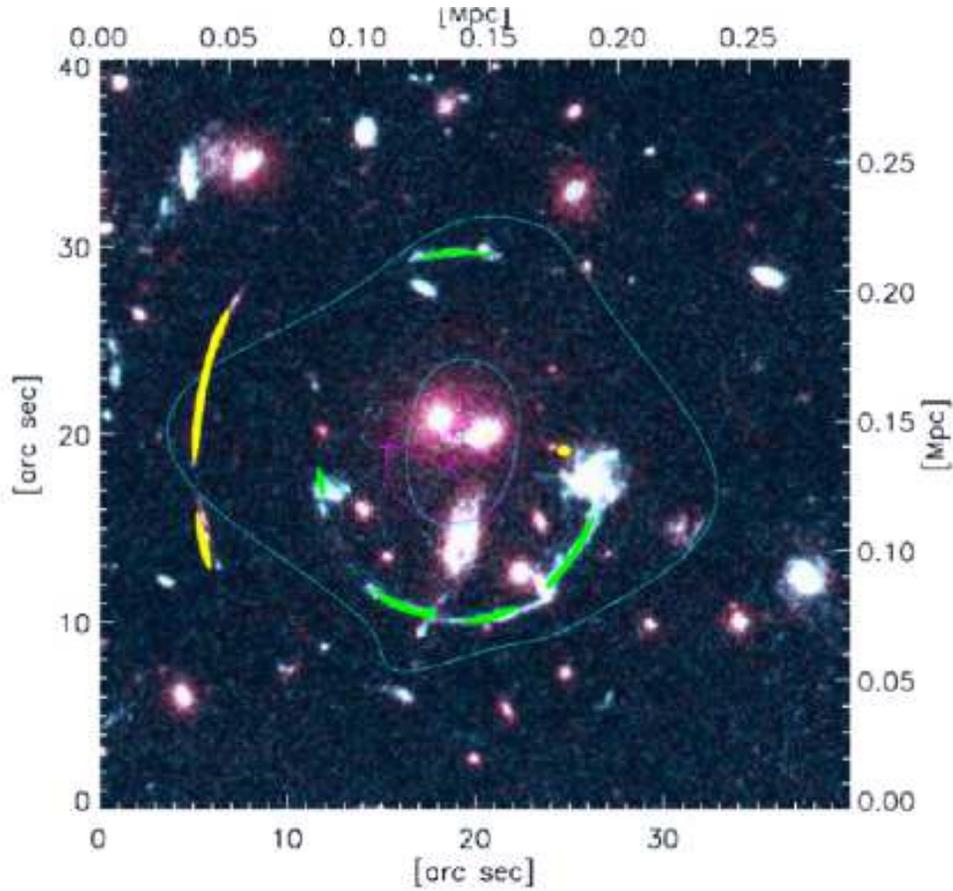}
   \caption{Images reproduced by our best fit model over-plotted on the combined F606W/F814W WFPC2 HST image. The closed lines show the critical curves and caustics for a source at $z=4.87$. The center of the image is at RA 02:24:34.218, Dec -00:02:31.64.}
   \label{fig.best_fit}
 \end{figure}

\begin{figure}[p]
   \centering
   \includegraphics[scale=0.6]{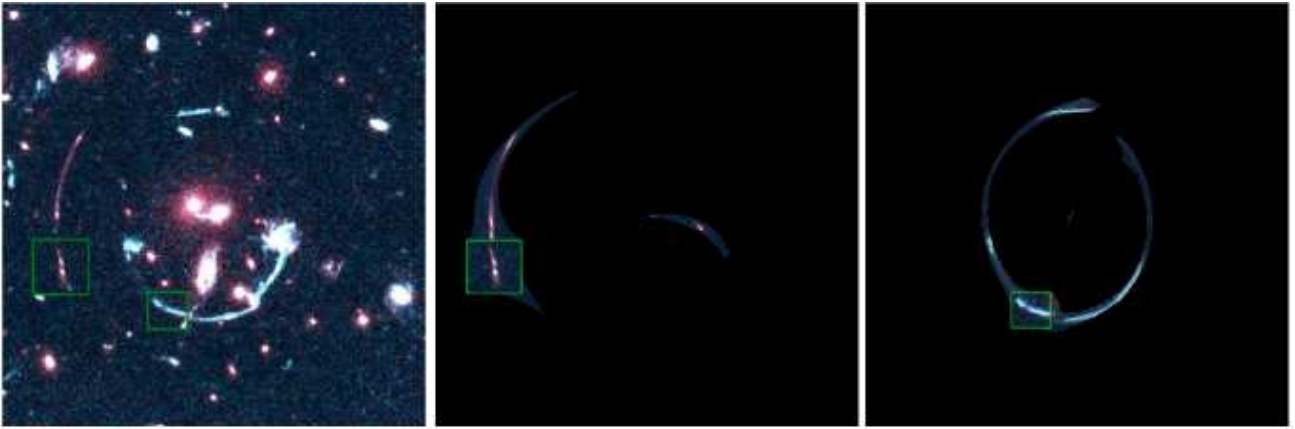}
   \caption{Result of image plane -- source plane -- image plane mapping. Panel to the left shows the arcs (marked by boxes) used to reproduce the arc systems. Middle and right panels show the arc systems as reproduced by the best fit mass model (the box marks the original image).}
   \label{lens-source-lens}
 \end{figure}

\begin{figure}[p]
   \centering
   \includegraphics[scale=0.6]{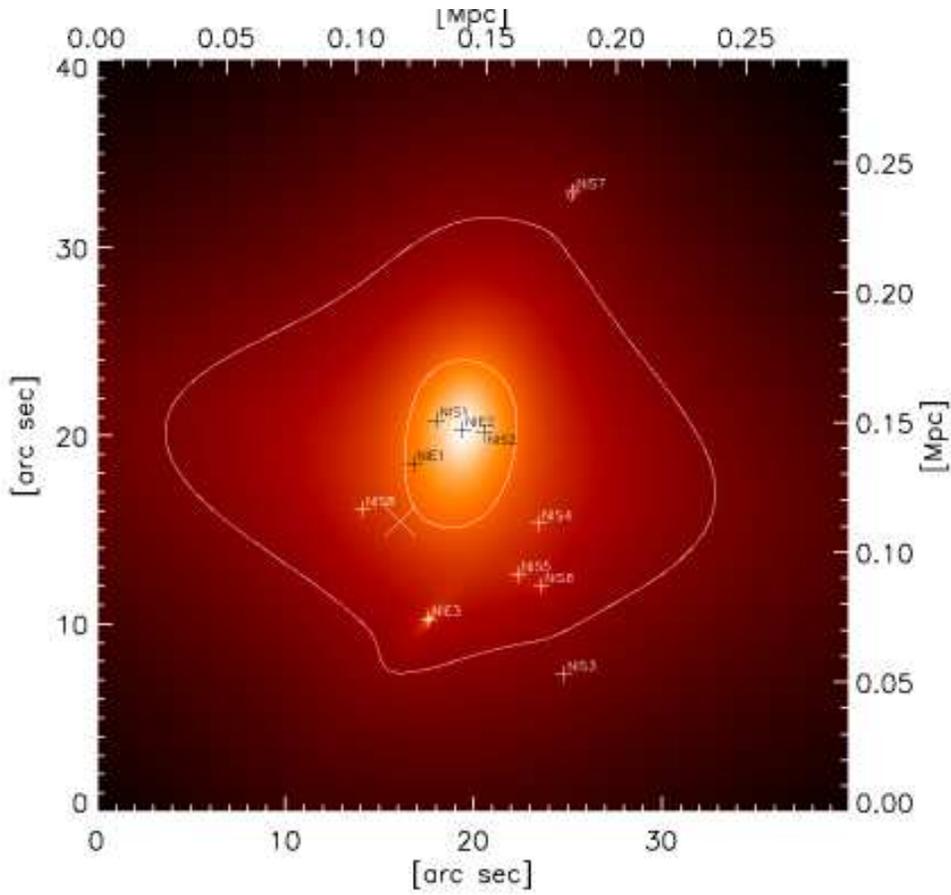}
   \caption{Mass density produced by our best fit model. The closed lines are the critical curves for a source at $z=4.87$. The crosses (+) mark the positions of our model components. The big cross (X) gives the position of the peak of the X-ray emission. The center of the image is at RA: 02:24:34.218 Dec: -00:02:31.64, the orientation as in Fig.~\ref{fig.xray}}
   \label{fig.smass}
 \end{figure}

\begin{figure}[p]
   \centering
   \includegraphics[scale=0.5]{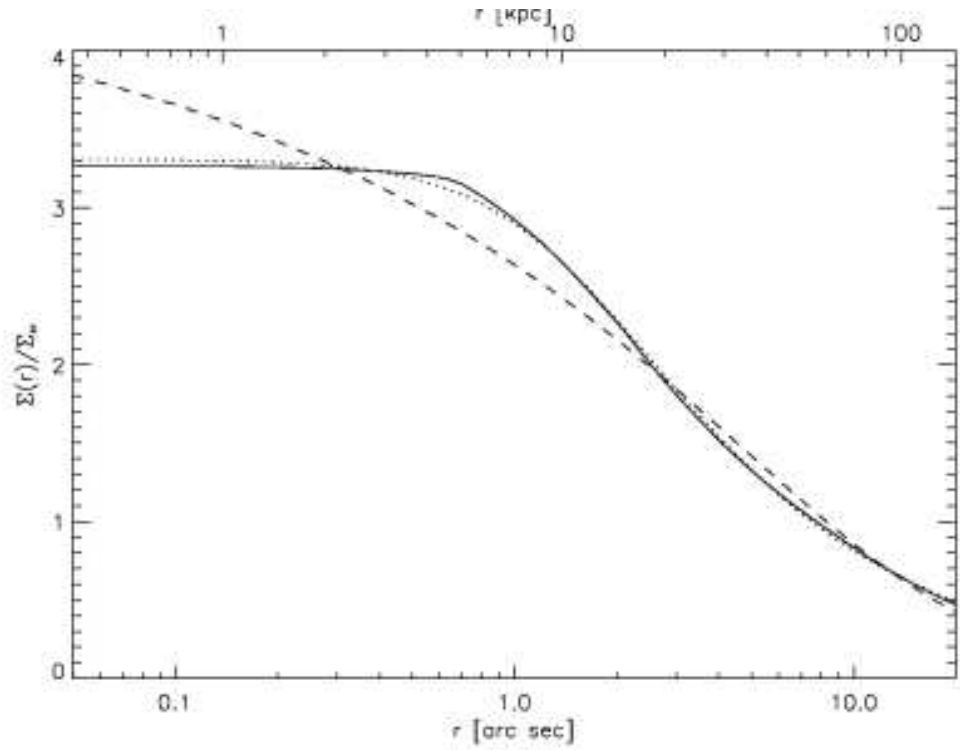}
   \caption{Radial average profile of the surface mass density of our best fit model (solid line) versus power law profile with $\gamma = 0.74$ (dotted line) and NFW-like profile with $\beta=0.69$ (dashed line).}
   \label{radial_profile}
 \end{figure}

The best
fit model (with MHD as defined by Eq.~(\ref{mhd}) equal to $30.3$) is
presented in Fig.~\ref{fig.best_fit}. The values of corresponding
parameters are given in Tables \ref{table:1_0} and \ref{table:1_1}.
The model reproduces fairly well all the observed strong lensing
features. The giant arc A include a counter-image 7\arcsec\ to the
west of the BCGs (A3). The model also reproduce the quadrupole system
B (B1,..B4). The central feature C is also predicted fairly close to
the observed one, although with different morphology. None of the
models we analyzed could reproduce the radial feature D, which
suggests that it is probably a foreground edge-on galaxy.  In
addition, inclusion of D to the lens model (NIE3) significantly
improved our fits and allowed us to ``break'' the arcs system B into
two arcs B1 and B3.  The best fit redshift of the source for the
system B is $2.65\pm 0.08$; a spectroscopic redshift of these blue
arcs, as well as object D, would provide a strong validation of our
lensing model and could also be used to better constrain the mass
distribution.  Estimates of the statistical errors are discussed in
the following section.  Figure~\ref{lens-source-lens} and
Tab.~\ref{stamps} show the results of some tests performed to assess
how well the best fit model is able to reproduce the morphology of
the multiple image systems A and B. For this purpose, we ray-trace a
given image for each system (A2 and B1, marked with green boxes in
Fig.~\ref{lens-source-lens}) into the source plane by using its HST
color image. This gives us the reconstructed source image.  We then
ray-trace back all the pixels from the source plane into the image
plane, thus finding all counter-images of the given image.  These
reconstructed counter-images are finally compared with the observed
ones (A1,3 and B2,3,4).  In general, we find a good agreement,
especially the knots in the A1 arc are very well reconstructed. The
overall shapes of all the arcs in the system B are also accurately
predicted.  The mass of the cluster within $R_{200} = 0.4
\mathrm{Mpc}$ obtained from the model is $1.9\pm 0.1\times 10^{14}\
\mathrm{M_\odot}$ and its distribution is shown in
Fig.~\ref{fig.smass}. This is in a good agreement with an mass
derived above from the X-ray temperature. Since we do not know all
the cluster member galaxies, we cannot reliably estimate the
mass-to-light ratio of the whole cluster. For the substructure 
(the mass associated with the luminous cluster component - NIS1..8), we
find an average mass-to-light ratio $\mathrm{M/L_{B,{\it vega}}}
\approx 3.6\ \mathrm{M_\odot/L_{\odot,B}}$. We converted the observed
F814W filter flux to the rest frame B filter flux, by calculating a
k-correction for a template elliptical galaxy from \citet{kinney}.
The center of the mass of the best fit model follows the light
distribution. NIE1 is found to be a diffuse (core radius $\approx
15\, \arcsec$) mass component close to the peak of the X-ray
emission.  The latter is shifted $\approx 5$'' from the NIE2
component, which corresponds to the center of the potential well and
the position of the BCGs. This may indicate the presence of a merger.
The radial average profile of the best fit surface mass density is
shown in Fig.~\ref{radial_profile}. This can be well approximated by
a power law profile with a slope $\gamma=0.74^{+0.03}_{-0.04}$, which
is closer to the isothermal profile ($\gamma=1$) than results
obtained in other clusters. For example, the analysis of the cluster
J1004-4112 yielded $\gamma\approx 0.5$ \citep{sharon2005} and $0.3 <
\gamma < 0.5$ \citep{williams2004}, whereas \citet{broadhurst} found
$\gamma = 0.5$ in A1689 using a large number of identified multiple
images. Note that the flat core of the mass profile we have found, being a
result of a hight value of the $r_c$ of the NIE1 component, is well
constrained by the position of the central arc C. The change of the
$r_c$ by 50\% causes the shift in the C arc position of $\approx 1$ arc sec.

By approximating the mass density distribution with NFW-like profile of
the form

\begin{align}
 \rho(r) ={}& \frac{\rho_0}{(r/r_c)^{\beta}(1+r/r_c)^{(1-\beta)}}\;,
\end{align}

we find a slope $0.69^{+0.09}_{-0.13}$, flatter then the canonical NFW
model ($\beta=1$), however in good agreement with other studies which
obtained $\beta < 1$. For example, \citet{sand} finds $\beta = 0.35$
for the galaxy cluster MS1237-23, and  $\beta < 0.57$ (at 99\%
confidence level) from the analysis of a large sample of clusters
\citep{sand2004}. 

\begin{figure}[p]
   \centering
   \includegraphics[scale=0.5]{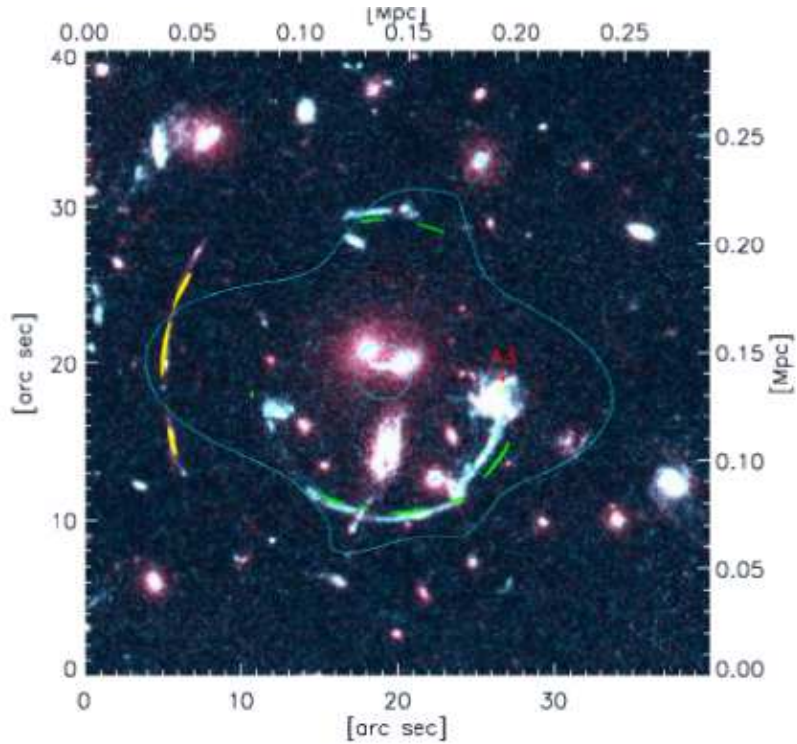}
   \caption{Images reproduced by our best fit NFW model over-plotted on the combined F606W/F814W WFPC2 HST image. The closed lines are the critical curves for a source at $z=4.87$. The center of the image is RA: 02:24:34.218 Dec: -00:02:31.64}
   \label{nfw_model}
 \end{figure}

In addition, we have tried to fit a model based the universal NFW
profile rather than NIE. The result, presented in the
Fig.~\ref{nfw_model}, shows that an NFW model performs significantly
worse then the NIE one.  The arcs A1 and A2 are reproduced fairly
well, but the counter image A3 is found much too far from the cluster
center. In addition, in the NFW model feature B4 is split into two
arcs (the second of which is not observed) and the reproduced arc B2
is shifted with respect to the observed one.  This is reflected by the
value of MHD, which is ten times bigger then the corresponding value
for the best-fit NIE model.  We note, however, that this bad
performance might be due to the approximated NFW elliptical model used
in our code, where the ellipticity is achieved by perturbing the
potential of the spherical NFW profile instead of its density.  This
approximation holds for potentials close to spherical, and therefore we
need to impose additional restrictions on the ellipticity of the NFW
components.

\section{Error analysis}
\label{error}

\begin{figure}[p]
   \centering
   \includegraphics[scale=1.0]{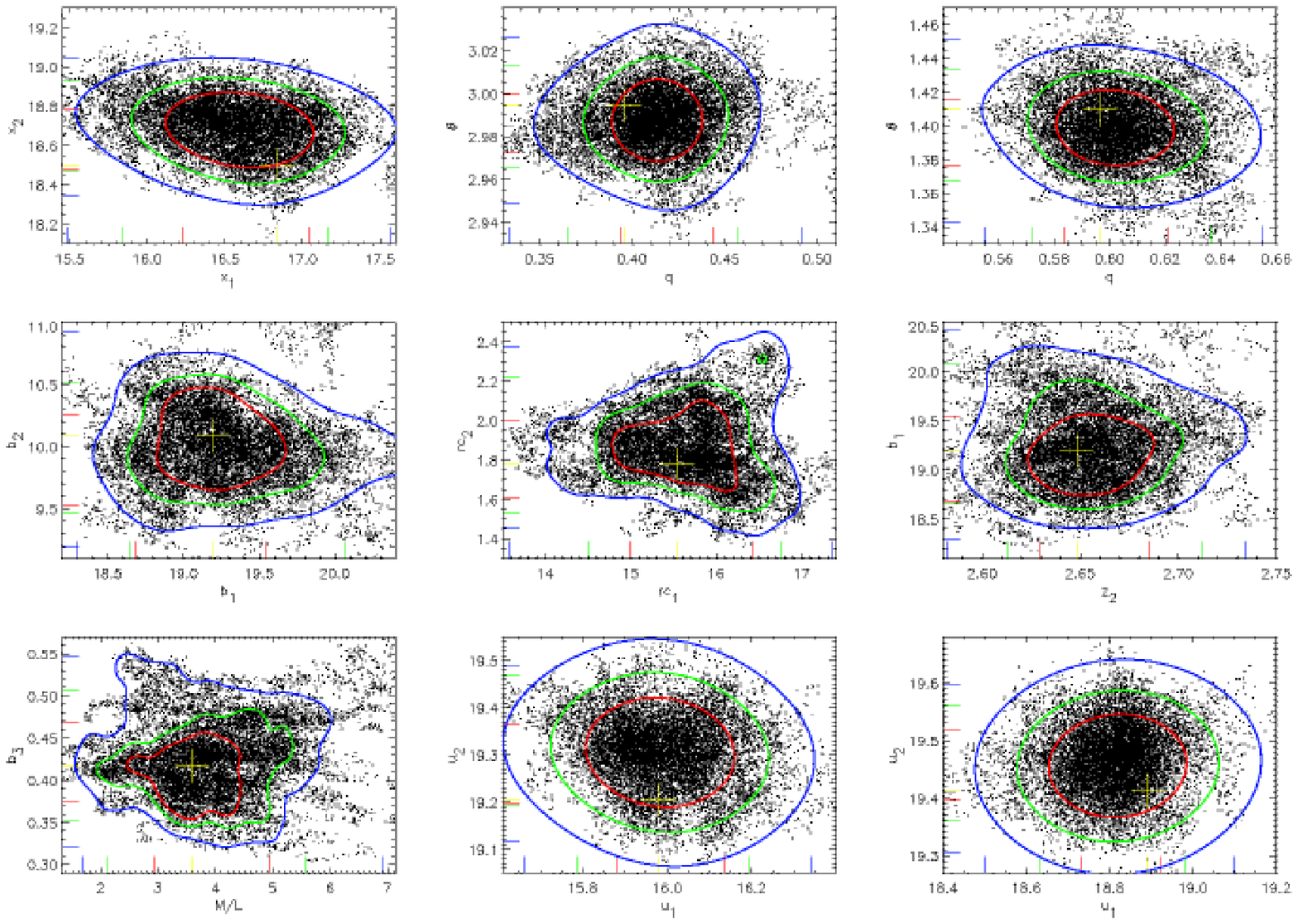}
   \caption{MCMC error estimates. The contours correspond to 68\%, 90\% and 99\% confidence levels. Marks on vertical and horizontal axis give the same confidence levels for 1D projected variables. The cross marks the position of the best fit point.}
   \label{mcmc_error}
 \end{figure}

Our method involves the minimization of the MHD whose expression
(Eq.18) is not a formal $\chi^2$ and includes a number of penalty
functions (weights) to limit the range of some parameters. As a result, it is
difficult to obtain reliable errors on the best fit parameters. In the
presence of many parameters, the Monte Carlo Markov Chain (MCMC, see
for example \citealp{neal}) method is an efficient way to estimate the
likelihood associated to our best fit model.  MCMC is used as a third
step of our minimization process by reconstructing the probability
distribution function of our model parameters. We start the
construction of Markov chain using the {\it Metropolis} algorithm
\citep{metropolis} from the best fit solution of the MHD
minimization. We use a number of chains randomly distributed around
the best fit point.  The resulting chain being the composition of all
those partial chains provides an approximate probability distribution
function for our parameters, from which we estimate the confidence
levels shown in the Fig.~\ref{mcmc_error}. Also by randomly probing
the parameters space, the MCMC algorithm helps to fine tune our best
fit parameters returned by the previous step of minimization. Most of
the parameters are well constrained (within 10 - 20 percent). The
unknown redshift of the arc system B appears to be well constrained,
$z_B=2.65\pm 0.08$. The mass to light ratio of the substructure is
however poorly constrained to be $3.6^{+3.3}_{-1.8}$.

\begin{figure}[p]
   \centering
   \includegraphics[scale=0.5]{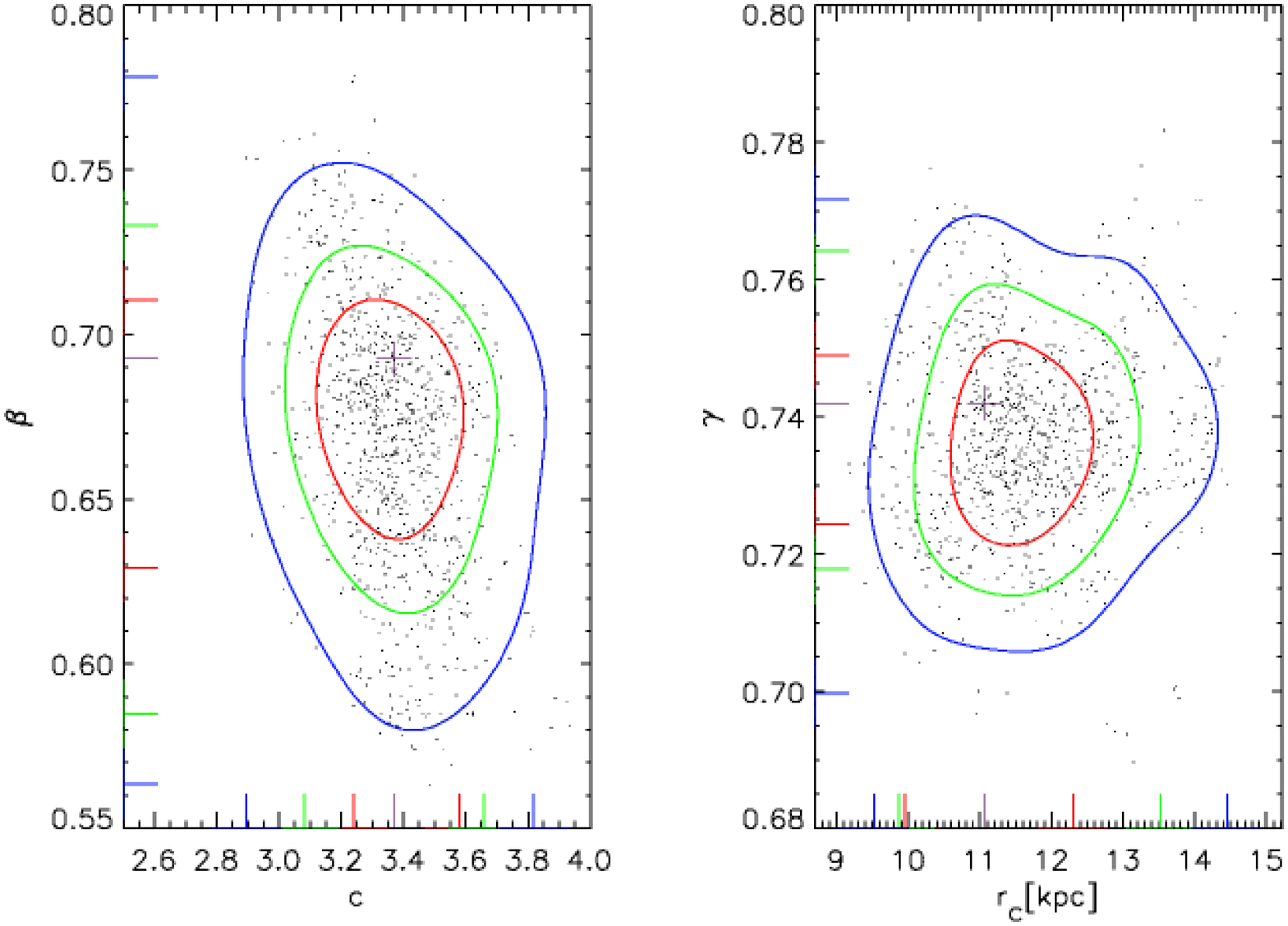}
   \caption{MCMC error estimates of the parameters of the single power law and NFW-like profiles fit. The contours correspond to 68\%, 90\% and 99\% confidence levels. Marks on vertical and horizontal axis give the same confidence levels for 1D projected variables. The cross marks the position of the best fit point.}
   \label{alpa_beta_mcmc}
 \end{figure}

We estimated the errors of a single power law and NFW-like profile
parameters by drawing a random sample of models from our Markov Chain,
and then fitting a single power law and NFW-like profile to that
sample. The resulting error estimates are presented in
Fig.~\ref{alpa_beta_mcmc}. This shows that isothermal and NFW
profiles are excluded with 99\% confidence level.

\section{Conclusions}
\label{conclusions}

We have performed a strong lensing analysis of the cluster
RCS0224-0002 using HST/WFPC2 images in F814W and F606W bands. We used
two arc systems: a red giant tangential arc 14\arcsec\, from the
center, with measured redshift of 4.87, for which we identified an inner
counter image, and a system of blue arcs at smaller radii with no
spectroscopic information.

We have modeled the mass distribution with with three mass
components: isothermal spheres associated with the most luminous
cluster members to model the substructure, and two isothermal
ellipsoids to model the underlying smooth mass component. Since
spectroscopic information is available in the literature only for two
cD galaxies, we identified likely member galaxies in the
cluster core from the red sequence, which is clearly detected in the
F606W-F814W color distribution. To infer the mass distribution from
the position and shapes of the strong lensing features we used a
three-step approach: i) minimization of the size of the two sources on
the source plane, ii) minimization of the difference between the
observed and modeled arcs on the image plane, based on the Modified
Hausdorff Distance, and iii) a refined estimate of the best fit parameters
and errors analysis with the Monte Carlo Markov Chain. The resulting mass density reproduces
all the strong features fairly well. The redshift of the blue arc
system is predicted to be $2.65 \pm 0.08$.

 We find that the substructure made of nine isothermal components centered on the
brightest cluster members, with
$\mathrm{M/L_{B,{\it vega}}} \approx 3.6 \mathrm{M_\odot/L_{\odot,B}}$
is crucial to exactly reproduce the shapes and positions of all the
arcs.

By fitting a single power-law or NFW-like halo to the radial average
mass density distribution we have found that both profiles are far
from canonical isothermal and standard NFW: we have found the
power-law parameter $\gamma$ to be $0.74^{+0.03}_{-0.04}$ ($\gamma =
1$ for an isothermal profile) and steepness parameter for NFW-like
profile $\beta$ to be $0.69^{+0.09}_{-0.13}$ ($\beta = 1$ for a NFW
profile), with the upper boundary very well constrained. Both those
values are consistent with the results obtained by studying the strong
lensing properties of other clusters \citep[see][]{sand, sand2004}.
The best fit NIS has $\sigma_v = 925\, \mathrm{km/s}$ and $r_c = 11\,
\mathrm{kpc}$; the best fit NFW has $R_{200} = 0.4 \mathrm{Mpc}$ and
concentration parameter $c = 3.4^{+0.4}_{-0.5}$, similarly to other
massive clusters ($c\approx 4$ for a z = 0.18 cluster \citealp{halkola}, 
$c\approx 5$ for z = 0.68 cluster \citealp{williams2004}). However, a wide range of
concentration parameters are found \citep[e.g. for $c>10$
see][]{broadhurst2005a}.  We have measured the total mass of the cluster
within $R_{200}$ to be $1.9\pm 0.1\times 10^{14}\, \mathrm{M_\odot}$
and its main component may be well described by a two NISs with a
$\sigma_{v1} = 945^{+30}_{-23}\, \mathrm{km/s}$, a $r_{c1} =
112^{+13}_{-14}\, \mathrm{kpc}$, a $\sigma_{v2} = 702^{+31}_{-28}\,
\mathrm{km/s}$, and a $r_{c2} = 12^{+4}_{-2}\, \mathrm{kpc}$.  The
mass of RCS0224-0002 derived from the lensing model is in a very good
agreement with the one obtained from the X-ray temperature measured
with deep Chandra observations ($M_{200}=(1.7\pm1.1)\times 10^{14} M_\odot$).

This analysis shows that even with a limited number of identified
multiple images we could constrain the mass distribution fairly
accurately. This was possible, in the case of RCS0224-0002, because
the two arcs systems are at very different angular diameter distances
and probe significant fraction ($\approx 20\%$ for the arcs system A, 
and $\approx 60\%$ for the system B) of the Einstein rings. Further
spectroscopic observations of the system B, as well as cluster
members, will allow a very robust constraint of the mass density
profile of the inner core of this cluster and its substructure.

\begin{acknowledgements}
  We would like to thank Matthias Bartelmann for very useful discussions
  and comments, specifically on the usage of the MCMC method to estimate
  the errors of our model.
\end{acknowledgements}

\bibliography{lens.bib}
\bibliographystyle{aa}

\onecolumn
\begin{table*}[t]
\caption{Parameters defining our model (see equation \ref{eqn.iso_ell:1}--\ref{eqn.iso_ell:3}) after  minimization. Parameters in parenthesis  were allowed to change during minimization}  
\label{table:1_0}
\centering
\begin{tabular}{c c c c c c c c c c c c} 
\hline\hline 
& NIE1 & NIE2 & NIS1 & NIS2 & NIS3& NIS4 & NIE3 & NIS5 & NIS6 & NIS7 & NIS8 \\
\hline
  $x_1$&  (16.834) & 19.413 & 18.039 & 20.578 & 24.799 & 23.494 & 17.621 & 22.389 & 23.614 & 25.296 & 14.097\\
  $x_2$&  (18.502) & 20.307 & 20.834 & 20.147 & 7.336  & 15.330 & 10.253 & 12.612 & 12.005 & 32.969 & 16.076\\
  $z$&  0.782 & 0.782 & 0.782 & 0.782 & 0.782 & 0.782 & 0.782 & 0.782 & 0.782 & 0.782 & 0.782 \\
  $b$&  (19.196) & (10.086) & (0.116)* & (0.191)* & (0.027)* & (0.032)* & (0.417) & (0.081)* & (0.075)* & (0.087)* & (0.037)* \\
  $q$&  (0.396) & (0.597) &  &  &  &  & 0.3  & & & & \\
  $\theta$& (2.994) & (1.409) &  & & & & 0.873 & & & &\\ 
   $s$& (15.539) & (1.778) & 0.1 & 0.1 & 0.1 & 0.1 & 0.1 & 0.1 & 0.1 & 0.1 & 0.1 \\ 
\hline
\end{tabular}\\
\footnotesize{$x_1,x_2$ : central position in arc seconds in the coordinate system of the Fig.~\ref{fig.smass},  $z$ : redshift,  $b$ :  scale factor in arc seconds,  $q$ : ellipticity, $\theta$ : position angle in radians, $s$ : core radius in arc seconds}\\
\footnotesize{* -- for the substructure the M/L ratio has been used as the variable for the minimization}\\
\end{table*}

\begin{table*}
\caption{Parameters defining sources after minimization. Parameters in parenthesis  were allowed to change during minimization}  
\label{table:1_1}
\centering
\begin{tabular}{c c c} 
\hline\hline 
& SOURCE1 & SOURCE2 \\
\hline
 $u_1$&  (15.979) & (18.892) \\
 $u_2$&  (19.204) & (19.412)  \\
 $z$&   4.878  & (2.648) \\
\hline
\end{tabular}\\
\footnotesize{$U_1$, $U_2$ : source position in arc seconds,  $z$ : redshift}\\
\end{table*}

\begin{table*}
\caption{Images reproduced under image plane -- source plane -- image plane mapping}  
\label{stamps}
\centering
\begin{tabular}{|c| c c c|} 
\hline\hline 
 \multirow{2}{*}{Image} & \multicolumn{3}{|c|}{Counter Images} \\
 & \multicolumn{3}{|c|}{Reproduced Images} \\
\hline
 A2 \includegraphics[scale=1.0]{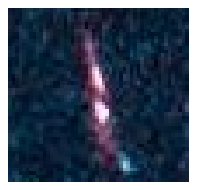} &  A1 & A3 & \\
\hline
 & \includegraphics[scale=1.0]{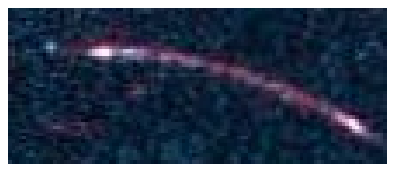} & \includegraphics[scale=1.0]{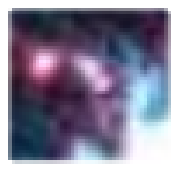} x2 &\\
& \includegraphics[scale=1.0]{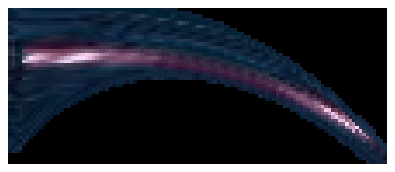} & \includegraphics[scale=1.0]{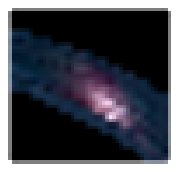} x2 &\\
\hline
B1 \includegraphics[scale=1.0]{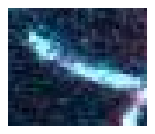} &  B2 & B3 & B4\\
\hline
& \includegraphics[scale=1.0]{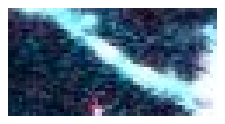} & \includegraphics[scale=1.0]{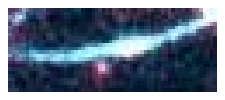}&\includegraphics[scale=1.0]{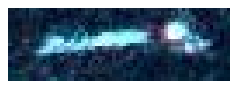}\\
& \includegraphics[scale=1.0]{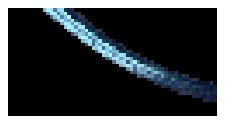} & \includegraphics[scale=1.0]{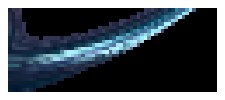}&\includegraphics[scale=1.0]{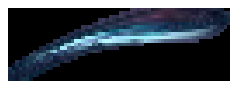}\\
\hline
\end{tabular}\\
\footnotesize{First column shows images used to construct sources. Second column shows both original and model reproduced images.}\\
\end{table*}

\end{document}